\documentclass[%
 reprint,
 amsmath,amssymb,
 aps,
]{revtex4-2}
\allowdisplaybreaks

\usepackage{color}
\usepackage{graphicx}
\usepackage{dcolumn}
\usepackage{multirow}
\usepackage{bm}

\begin{document}

\preprint{APS/123-QED}

\title{Distinguishing limit of Bell states for any $n$-photon $D$-dimensional hyperentanglement}

\author{Chunzhen Li}
\email{These authors contributed equally to this work.}
\author{Yi Li}
\email{These authors contributed equally to this work.}
\affiliation{Key Laboratory of Weak-Light Nonlinear Photonics and School of Physics, Nankai University, Tianjin 300071, China}

\author{Yongnan Li }
\email[Corresponding author. Email: ]{liyongnan@nankai.edu.cn}
\affiliation{Key Laboratory of Weak-Light Nonlinear Photonics and School of Physics, Nankai University, Tianjin 300071, China}

\date{\today}

\begin{abstract}
Bell state measurement is crucial to quantum information protocols, but it is impossible to unambiguously distinguish all the Bell states encoded in multi-photon using only linear optics. There is a maximum number of distinguished Bell states, i.e. distinguising limit which is very important for increasing the channel capacity of quantum communications. In this paper, we separate  $n$-photon $D$-dimensional hyperentanglement into two groups. For the first group of $U$ ($G_1$), we obtain the limit ${N_1} = nD - (n - 1)$, which can be applied for both bosons' and fermions' cases. We further discuss the limit $N$ for any $nD$ system with the second group of $U$ ($G_2$), inferring that at least ${D^{n - 1}}$ Bell states can be distinguished due to the symmetry of Bell states. Obviously, ${N_1} \le {N_2}$ for those systems with $n>2$. Finally, we theoretically design an optical setup for Bell state measurement of two-photon eight-dimensional hyperentanglement of spin, path and orbital angular momentum (OAM) and distinguish 15 classes of 64 Bell states. Our results provide a theoretical basis and practical reference to increase the channel capacity of the quantum communication.

\end{abstract}

\maketitle


\section{\label{sec:level1}INTRODUCTION}
Bell state measurement (BSM) is a crucial measurement in quantum mechanics, using joint orthogonal projection measurement. It enables many applications in quantum information processing, such as superdense coding (SDC) \cite{1,2}, teleportation \cite{A,B,C}, entanglement swapping \cite{D,E} and quantum finger printing \cite{F,G}. An important parameter of the quantum communication is a channel capacity (CC), which is determined by the number of distinguished Bell states by BSM  \cite{3}. However, a complete BSM with linear optics is impossible \cite{4,5} and its optimal probability of success is only 50\% \cite{5,6,7}.  In other words, there is an upper bound of CC for any given entangled system. As a result, the rate of quantum information between a sender (Alice) and a receiver (Bob) is limited. Many attempts have been made to establish optimal BSM schemes and achieve their limits using high-dimensional entanglement in both theory and experiments \cite{13,19,20,22}.

Actually, high-dimensional entanglement has many advantages over two-qubit entanglement, such as higher CC \cite{8,9} and transmission rate \cite{10,11}. It is indeed effective to completely distinguish more than four Bell state using hybrid degrees of freedom (DOFs), such as path \cite{13} and orbital angular momentum (OAM) \cite{14},  which is called hyperentanglement. Recently, other similar BSM schemes of hyperentanglement have been demonstrated in various optical systems \cite{15,16,17,18}. Some special cases were proved in theory and demonstrated by experiments. Wei \emph{et~al}. theoretically presented that the optimal scheme divides 16 Bell states into 7 distinct groups \cite{19}. Pisenti \emph{et~al}. showed that at most $2^{n+1}-1$ classes out of $4^{n}$ hyperentangled Bell states can be distinguished with one copy of input states \cite{20}. Hu \emph{et~al}. gave a four-dimensional entanglement protocol, which exceeded the channel capacity limit of 2 for the first time \cite{13}. Kong \emph{et~al}. produced eight eight-dimensional Bell-like states and distinguished them completely \cite{22}. This is because adding additional DOFs also enlarges Hilbert space and hence the number of Bell-like states also increases. However it is interesting and necessary for practice that what is the theoretical limit for distinguished Bell states in Hilbert space.

In our paper, we firstly investigate the distinguishing limit $N$ of Bell states for any $n$-photon $D$-dimensional hyperentangled system ($nD$ system). To achieve this, for all kinds of unitary transformation $U$ in the input-output relation $\vec c  = U\vec a$ for any linear-optics circuits, we separate them into two groups using a norm ($G_1$ and $G_2$) \cite{20}.   With the help of Pisenti’s work \cite{20}, we extend the method to $n$-photon systems and find that the limit for $G_1$ is ${N_1} = nD - (n - 1)$. We also verify this result using Peter van Loock and Norbert L\"{u}tkenhaus's criterion (LL criterion) \cite{21}. This limit can not only apply to bosons’ entanglement, but fermions’ as well. For the second group of $U$ ($G_2$) we infer that at least $D^{n - 1}$ Bell states can be distinguished for any $nD$ system due to the symmetry of Bell states and the limit for $G_2$ is ${N_2} \ge {D^{n - 1}}$. The BSM scheme for a four-photon two-dimensional system with $G_2=E$ ($E$ is the identity matrix) has been illustrated. It is clearly that for any entangled system with $n>2$, ${N_1} \le {N_2}$. Finally, we report an experimental demonstration theoretically for two-photon eight-dimensional hyperentanglement using spin, path and the first order of OAM, based on Kong's \cite{22} and Hu's \cite{13} work. 64 Bell states can be divided into 15 different classes, which achieves CC's upper limit ${\log _2}15$ for this case. 

\section{\label{sec:level2}THEORY}
The $D$-dimensional Bell basis of an $n$-particle system can be written as \cite{8}:

\begin{equation}
\begin{aligned}
\left| {\varphi _{{i_1},{i_2},...,{i_{n-1}}}^P} \right\rangle  =& \frac{1}{{\sqrt D }}\mathop \sum \limits_{j = 0}^{D - 1} {e^{i\frac{{2\pi }}{D}jP}}\left| j \right\rangle \left| {(j + {i_1})\bmod D} \right\rangle  \otimes\\
&... \otimes \left| {(j + {i_{n-1}})\bmod D} \right\rangle ,\label{eq3}
\end{aligned}
\end{equation}
where $P,j,{i_1},{i_2},...,{i_{n-1}} = 0,1,...,D - 1$. We define the vectors $\vec a=(\hat a_{{{1}}}^\dag,\hat a_{{{2}}}^\dag,\cdots,\hat a_{{{nD}}}^\dag)^T$ and $\vec c=(\hat c_{{{1}}}^\dag,\hat c_{{{2}}}^\dag,\cdots,\hat c_{{{nD}}}^\dag)^T$, where $\hat a_{{{i}}}^\dag$ and $\hat c_{{{i}}}^\dag$ represent an input and an output mode separately. Then linear-optics devices functioning as the unitary matrix $U$ in the circuits transform $\vec a$ to $\vec c$ follow the input-output relation $\vec c  = U\vec a$. If the arbitrary initial $n$-particle Bell state is $\left| {{\varphi _j}} \right\rangle$, the state following $n-1$ clicks in some detectors is proportional to ${\hat c_{{s_{n-1}}}} \cdots {\hat c_{{s_2}}}{\hat c_{{s_1}}}\left| {{\varphi _j}} \right\rangle$, where $\hat c_{{{s_i}}}$ ($s_i = 1,2,\ldots,nD$) is the annihilation operator associated with corresponding output mode. We call these states post-click states. Then we can separate all kinds of $U$ into two groups, $G_1$ and $G_2$, by the result of the norm of the post-click states with $n-1$ clicks $\left\langle {{\varphi _j}} \right|\hat c_{{s_1}}^\dag \hat c_{{s_2}}^\dag  \cdots \hat c_{{s_{n - 1}}}^\dag {\hat c_{{s_{n - 1}}}} \cdots {\hat c_{{s_2}}}{\hat c_{{s_1}}}\left| {{\varphi _j}} \right\rangle$. The norm shows the probability of $n-1$ clicks in the corresponding detectors, where we can easily find that the norm is nonnegative.

To define $G_1$, the norm for any combination of ${\hat c_{{s_i}}}$ and $\left| {{\varphi _j}} \right\rangle$ is nonzero. At this time, all the unitary matrices $U$ in $G_1$ must satisfy Eq.\ \eqref{eq13}

\begin{equation} \label{eq13} 
\setlength{\abovedisplayskip}{1pt}
\begin{split}
&\left\langle {{\varphi _j}} \right|\hat c_{{s_1}}^\dag \hat c_{{s_2}}^\dag  \cdots \hat c_{{s_{n - 1}}}^\dag {\hat c_{{s_{n - 1}}}} \cdots {\hat c_{{s_2}}}{\hat c_{{s_1}}}\left| {{\varphi _j}} \right\rangle \\
&=\sum\limits_{{p_1=1}}^{nD}\sum\limits_{{q_1=1}}^{nD}\cdots\sum\limits_{{p_{n-1}=1}}^{nD}\sum\limits_{{q_{n-1}=1}}^{nD} U_{{s_1}{p_1}} \cdots U_{{s_{n - 1}}{p_{n - 1}}}{U^*_{{s_{n - 1}}{q_{n - 1}}}} \\
& \cdots {U^*_{{s_1}{q_1}}} \left\langle {{\varphi _j}} \right|\hat a_{{p_1}}^\dag \hat a_{{p_2}}^\dag  \cdots \hat a_{{p_{n - 1}}}^\dag {{\hat a}_{{q_{n - 1}}}} \cdots {{\hat a}_{{q_2}}}{{\hat a}_{{q_1}}}\left| {{\varphi _j}} \right\rangle \\
&\neq 0,
\end{split}
\end{equation}
where when ${{\hat a}_{{p_{1}}}}{{\hat a}_{{p_{2}}}} \cdots {{\hat a}_{{p_{n-1}}}}={{\hat a}_{{q_{1}}}}{{\hat a}_{{q_{2}}}} \cdots {{\hat a}_{{q_{n-1}}}}$, $\left\langle {{\varphi _j}} \right|\hat a_{{p_1}}^\dag \hat a_{{p_2}}^\dag  \cdots \hat a_{{p_{n - 1}}}^\dag {{\hat a}_{{q_{n - 1}}}} \cdots {{\hat a}_{{q_2}}}{{\hat a}_{{q_1}}}\left| {{\varphi _j}} \right\rangle$ can be nonzero. This nonzero result means that via $G_1$, every Bell state is capable of producing $n-1$ clicks in all detectors randomly, but these detector clicks provide nothing about which Bell state particles occupy. In other words, the last click will determine the distinguishing of Bell states. Thus, we can get the distinguishing limit $N_1$ by analysing all the Bell state's last click to simplify the problem.

To define $G_2$, on the contrary, the norm for some combinations of ${\hat c_{{s_i}}}$ and $\left| {{\varphi _j}} \right\rangle$ is zero. For instance, identity matrices belong to $G_2$. In other words, when $\vec c = \vec a$, we can easily derive from Eq.\ \eqref{eq3} that the norm $\left\langle {{\varphi _j}} \right|\hat a_{{s_1}}^\dag \hat a_{{s_2}}^\dag  \cdots \hat a_{{s_{n - 1}}}^\dag {\hat a_{{s_{n - 1}}}} \cdots {\hat a_{{s_2}}}{\hat a_{{s_1}}}\left| {{\varphi _j}} \right\rangle$ for some combination of ${\hat a_{{s_i}}}$ and $\left| {{\varphi _j}} \right\rangle$ is zero. Thus for $G_2$ we cannot simplify the problem as the case of $G_1$ and more discussion on $G_2$ is shown at the end of this section.

With $G_1$, we can derive $N_1$ for any $nD$ system by generalizing Pisenti's method for two-photon systems \cite{20}.  
Firstly, we define every output mode $\hat c_{{i}}^\dag$ $(i=1,2,\ldots,nD)$ corresponds to a detector mode $\left| {{D_i}} \right\rangle$, which takes the form as follows    
\begin{equation} \label{eq12}
    \left| {{D_i}} \right\rangle  = ({\alpha _{{i_1}}}{\hat b^\dag }_{{i_1}} + {\alpha _{{i_2}}}{\hat b^\dag }_{{i_2}} + \cdots  + {\alpha _{{i_n}}}{\hat b^\dag }_{{i_n}})\left| 0 \right\rangle ,
\end{equation}
where ${\hat b^\dag }_{{i_j}}$ must come from $n$ different particles and ${\hat b^\dag }_{{i_j}}\left| 0 \right\rangle$ is a superposition of the $j^{\text{th}}$-particle's input state. We define the detection signature as ${P_{12 \ldots n}}\left| {{D_{{i_1}}}} \right\rangle \left| {{D_{{i_2}}}} \right\rangle  \ldots \left| {{D_{{i_n}}}} \right\rangle$ \cite{20}. The projection operator ${P_{12 \ldots n}}$ projects the raw tensor product of $n$ detector modes $\left| {{D_{{i_1}}}} \right\rangle \left| {{D_{{i_2}}}} \right\rangle  \ldots \left| {{D_{{i_n}}}} \right\rangle$ onto the subspace of $n$-particle states, which makes sure the modes come from different particles. Since every Bell state can be represented by an independent detection signature (IDS), the number of IDS equals to that of distinguishable Bell states.


For two-particle $D$-dimensional hyperentangled systems, the norm Eq.\ \eqref{eq13} is always positive \cite{20}. All kinds of $U$ are in $G_1$ here, thus there are $2D$ detection signatures with the post-click states with one click. Following Pisenti's method \cite{20}, the number of IDS is $2D-1$ for both bosons and fermions cases. 
When we just consider ${m}$ qubitlike degrees of freedom on each photon, $D = {2^{{m}}}$ and ${N_1}= {2^{{m} + 1}} - 1$, which is just the same as Pisenti's result \cite{20}.

When we extend Pisenti's method \cite{20} to $nD$ systems with $G_1$ and consider the post-click states with $n-1$ clicks, there are $nD$ detection signatures. For bosons with three-photon $D$-dimensional entanglement, there are $3D$ detection signatures if considering the post-click states with the first two clicks in $D_i$ as $\left| {{D_i}} \right\rangle \left| {{D_i}} \right\rangle=({\alpha _{{i_1}}}{{\hat b}^\dag }_{{i_1}} + {\alpha _{{i_2}}}{{\hat b}^\dag }_{{i_2}} + {\alpha _{{i_3}}}{{\hat b}^\dag }_{{i_3}})^2\left| 00 \right\rangle$. 
However, these $3D$ detection signatures are not independent. Next, we will show that there are only two independent constraint equations among them. 
Firstly, we can always find some linear combination of detector modes satisfy
\begin{equation} \label{eq14}
\begin{aligned}
\left| {{X}} \right\rangle  = \sum\limits_j {{\varepsilon _j}} \left| {{D_j}} \right\rangle  = A({\alpha _{{i_1}}}{{\hat b}^\dag }_{{i_1}} + B{\alpha _{{i_2}}}{{\hat b}^\dag }_{{i_2}} + C{\alpha _{{i_3}}}{{\hat b}^\dag }_{{i_3}})\left| 0 \right\rangle ,
\end{aligned}
\end{equation}
then we have ${P_{123}}\left| {{D_i}} \right\rangle \left| {{D_i}} \right\rangle \left| {{X}} \right\rangle = 0$ if and only if $B+C = -1$. For example,
\begin{equation} \label{eq142}
\begin{aligned}
\left| {{X_1}} \right\rangle  = \sum\limits_j {{\varepsilon _j}} \left| {{D_j}} \right\rangle  = A({\alpha _{{i_1}}}{{\hat b}^\dag }_{{i_1}} - \frac{1}{2}{\alpha _{{i_2}}}{{\hat b}^\dag }_{{i_2}} - \frac{1}{2}{\alpha _{{i_3}}}{{\hat b}^\dag }_{{i_3}})\left| 0 \right\rangle ,\\
\left| {{X_2}} \right\rangle  = \sum\limits_j {\varepsilon _j^{'}} \left| {{D_j}} \right\rangle  = A({\alpha _{{i_1}}}{{\hat b}^\dag }_{{i_1}} - \frac{1}{3}{\alpha _{{i_2}}}{{\hat b}^\dag }_{{i_2}} - \frac{2}{3}{\alpha _{{i_3}}}{{\hat b}^\dag }_{{i_3}})\left| 0 \right\rangle .
\end{aligned}
\end{equation}
In the three-dimensional Vector space, $\vec v_1=(1,-\frac{1}{2},-\frac{1}{2})$ and $\vec v_2=(1,-\frac{1}{3},-\frac{2}{3})$ are linearly independent. Since $\vec v_3=(1,B,-1-B)$ is on the plane made up of $\vec v_1$ and $\vec v_2$, $\vec v_3$ can be represented by $\vec v_1$ and $\vec v_2$ for any constant $B$. In other words, any $\left| X \right\rangle$ satisfying ${P_{123}}\left| {{D_{{i}}}} \right\rangle \left| {{D_{{i}}}} \right\rangle \left| {{X}} \right\rangle = 0$ can be represented by $\left| {{X_1}} \right\rangle$ and $\left| {{X_2}} \right\rangle$. Thus there are only two independent constraint equations, and they are
\begin{equation} \label{eq15}
\begin{aligned}
\sum\limits_j {{\varepsilon _j}} {P_{123}}\left| {{D_{{i}}}} \right\rangle \left| {{D_{{i}}}} \right\rangle \left| {{D_j}} \right\rangle {\rm{ = }}0,\\
\sum\limits_j {\varepsilon _j^{'}} {P_{123}}\left| {{D_{{i}}}} \right\rangle \left| {{D_{{i}}}} \right\rangle \left| {{D_j}} \right\rangle {\rm{ = }}0.
\end{aligned}
\end{equation}
The last click in detectors matters to distinguish Bell states with $G_1$, whether the first two detector modes are the same or not. Therefore, the number of IDS is at most $3D-2$. Similarly, there are $nD$ detection signatures for $nD$ systems with post-click states with $n-1$ detector clicks. 
But we can always find $n-1$ independent constraint equations, so the number of IDS in this case is 
\begin{equation} \label{eq7}
    N_1=nD-(n-1).
\end{equation}
Here, we note that we cannot completely distinguish $N_1$ Bell states of $nD$ system without number-resolving detectors.

For Fermions, anti-symmetry under particle exchange is a must. Thus, $n$ clicks must trigger $n$ different detectors to get allowed post-click states. 
For the post-click states with $n-1$ different clicks, the last click has only $nD-(n-1)$ possibilities. That is to say, there are at most $nD-(n-1)$ detection signatures as well as the number of IDS.

We also verify this result by LL criterion. It shows that the set of necessary and sufficient conditions for complete discrimination between two states $\left| {{\varphi _i}} \right\rangle $ and $\left| {{\varphi _j}} \right\rangle $ $(i \ne j)$ with linear optics are \cite{21}:
\begin{equation}
\begin{aligned}
\left\langle {{\varphi _i}} \right|{ {{{\hat c_s}^\dag }} } {\hat c_s} \left| {{\varphi _j}} \right\rangle  &= 0,\qquad \forall s,\\
\left\langle {{\varphi _i}} \right|{ {{{\hat c_s}^\dag }{{{\hat c_{s^{'}}}^\dag} }} } { {{\hat c_s}{{\hat c_{s^{'}}} }} } \left| {{\varphi _j}} \right\rangle  &= 0,\qquad  \forall s,\forall s^{'},\\
\left\langle {{\varphi _i}} \right|{ {{{\hat c_s}^\dag }{{{\hat c_{s^{'}}}^\dag} }} } { {{{\hat c_{s^{''}}}^\dag }{{\hat c_s} }} }{ {{\hat c_{s^{'}} }{{\hat c_{s^{''}}} }} } \left| {{\varphi _j}} \right\rangle  &= 0,\qquad  \forall s,\forall s^{'},\forall s^{''}.\\
\vdots \qquad & = \qquad \qquad \vdots
\label{eq1}
\end{aligned}
\end{equation}


Because the norm Eq.\ \eqref{eq13} is nonzero with $G_1$, different output mode operators in Eq.\ \eqref{eq1} can be all substituted with a particular mode operator ${\hat c_s}$, shown as Eq.\ \eqref{eq2},

\begin{equation}
\left\langle {{\psi _i}\left| {{\psi _j}} \right\rangle } \right. = 0,{\rm{\;\;\;\;\;\;\;\;\;}}\left| {{\psi _j}} \right\rangle  = {\left( {\hat c_s} \right)^m}\left| {{\varphi _j}} \right\rangle, \forall m \ge 1. \label{eq2}
\end{equation}
Actually, using Eq.\ \eqref{eq2} with $m=n-1$, we can simply derive the result ${N_1}=nD-(n-1)$ by summarizing the results of many $nD$ systems. Here we show the distinguishing limit of some specific systems with the help of LL criterion in TABLE \ref{t1}.

\begin{table}[ht]
	\centering
	\begin{tabular}{|c|c|c|}\hline
		Particle Number $n$&Dimension $D$&the Limit $N_1$\\\hline
	\multirow{3}{*}{3}&2&4\\\cline{2-3}
	&3&7\\\cline{2-3}
	&4&10\\\hline
	\multirow{3}{*}{4}&2&5\\\cline{2-3}
	&3&9\\\cline{2-3}
	&4&13\\\hline
	\end{tabular}
	\caption{\textbf{distinguishing limit of some specific $nD$ systems calculated by LL criterion.}}
	\label{t1}
\end{table}



\begin{figure}[htb]
\includegraphics[width=0.45\textwidth]{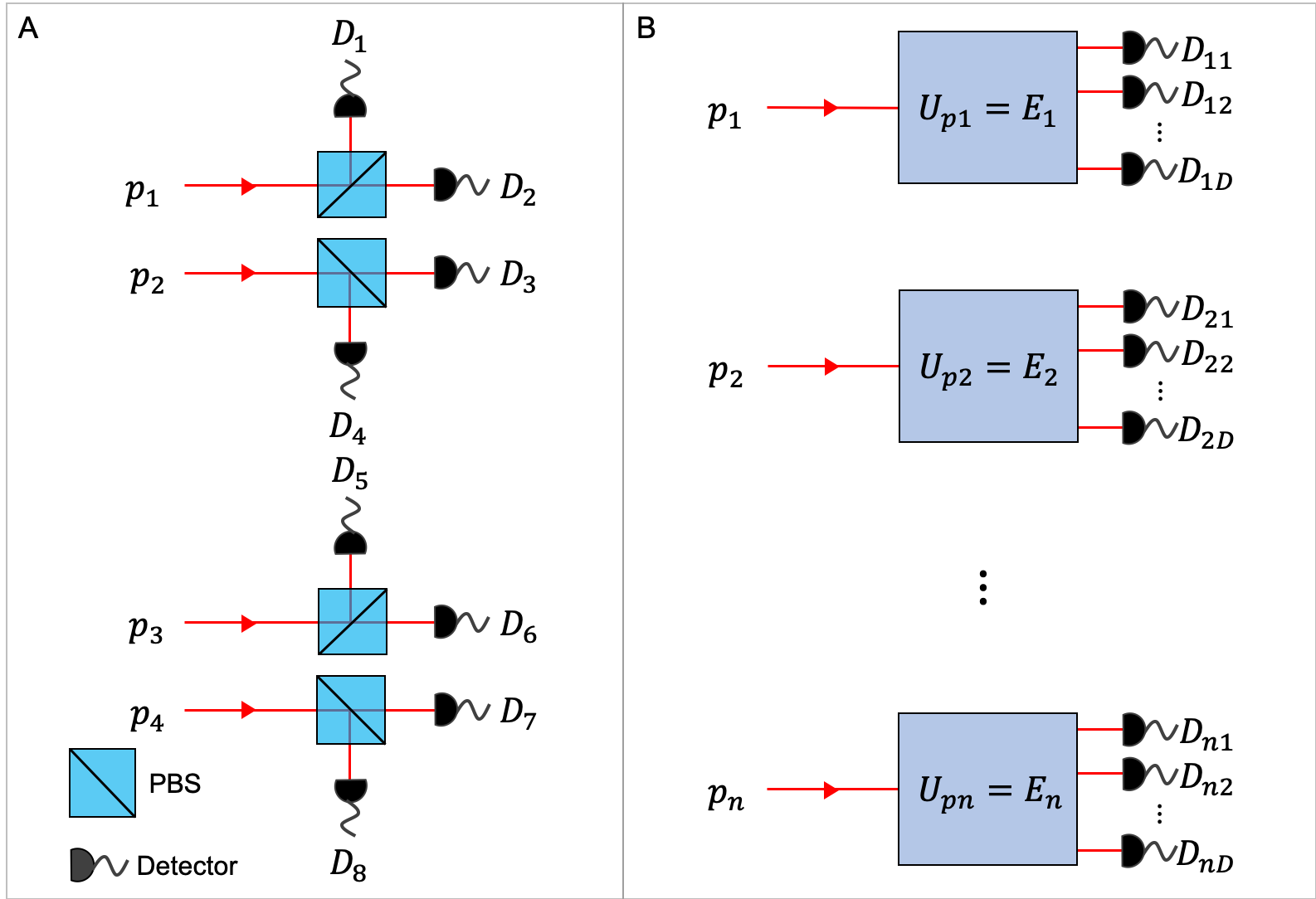}
\caption{\label{figA} \textbf{BSM schemes with $U2=E$ to distinguish Bell states.} $p_i$ represents a photon (A) Scheme for a four-photon two-dimensional system using spin. Every photon has chances to reach two detectors. 8 Bell states can be distinguished completely. (B) Scheme for any $nD$ system. ${U_{{p_i}}}$ represents a single-particle unitary transformation and $E_i$ represents the corresponding identity matrix. Every photon has chances to reach $D$ detectors. $D^{n-1}$ Bell states can be distinguished completely.}
\end{figure}

When it comes to $G_2$, the number of detection signatures cannot be calculated simply and Eq.\ \eqref{eq2} is not a necessary and sufficient condition any more. But we can still easily distinguish $D^{n-1}$ Bell states with $U=E$ due to Bell-state symmetry. As shown in Fig.\ref{figA}, for the four-photon two-dimensional BSM scheme using spin, each PBS act as an identity matrix $E$ to transform the input mode ${{\hat a}^\dag }_i$ to the detector mode $\left| {{D_i}} \right\rangle$. We can easily distinguish $8$ Bell states due to the symmetry. For the $nD$ system, we can similarly use some linear-optics devices functioning as $U_{p_i}=E_i$ like PBS. Therefore, $D^{n-1}$ Bell states can be distinguished completely with $G_2=E$ due to the symmetry.

Comparing with the case of $G_1$ mentioned above, it is clearly that $N_1\leq N_2$ for any entangled system with $n>2$. The equality sign holds if and only if $N_2=D^{n-1}$, $n=3$ and $D=2$. In addition, in two-photon systems measurement efficiency ${\rm{ME}} = \frac{{2D - 1}}{{{D^2}}}$, and in multi-photon systems ${\rm{ME}} \ge \frac{1}{D}$. Except two-photon two-dimensional systems, all the other two-photon systems' ME is less than $\frac{1}{2}$. While for multi-photon two-dimensional systems, ME is at least $\frac{1}{2}$. Thus, we can come to the conclusion that high-dimensional multi-photon hyperentanglement with $G_2$ circuits has great potential to achieve high CC and ME at the same time.


\begin{figure}
\includegraphics[width=0.45\textwidth]{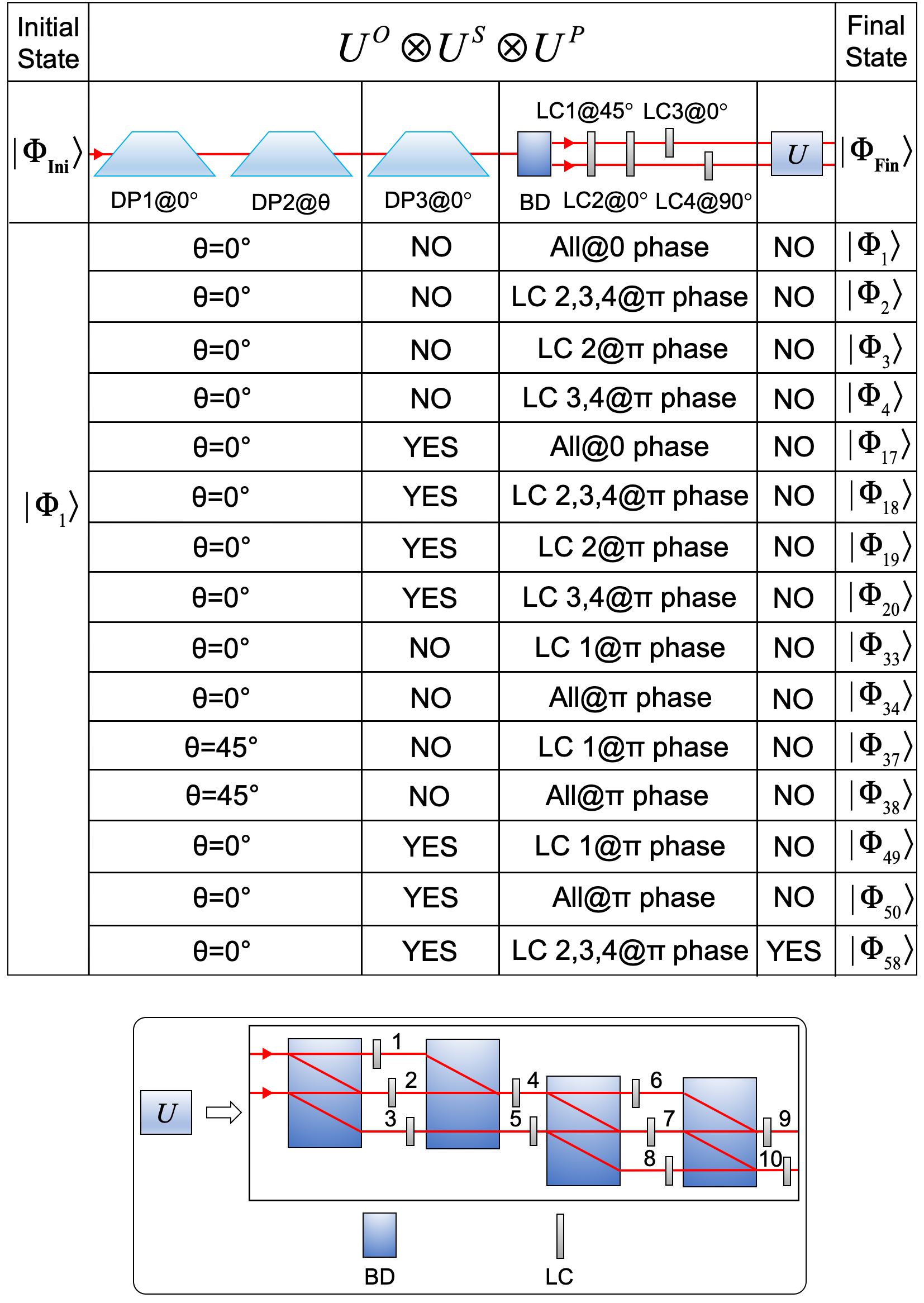}
\caption{\label{fig2} \textbf{Transformation for 15 Bell-like states.} Preparation of Bell-like states from $\left| {{\Phi _1}} \right\rangle $ by manipulating unitary operation ${U^O} \otimes {U^S} \otimes {U^P}$ on photon A. Here ${U^O}$ is based on Kong’s work \cite{22} and is composed of three Dove prisms (DPs). $\theta$ is the relative orientation angle between DP1 and DP2. To realize the conversion of the $1^{\text{th}}$-order OAM Bell states, $\theta$ should be set at 45 degree. ${U^S} \otimes {U^P}$ is based on Hu’s work \cite{13} and composed of one beam displacer (BD) and four computer-controlled liquid crystal variable retarders (LCs). The optical axes of the LCs are set at different angles. By applying different voltages, the LCs will introduce different phases between the fast and slow axis, leading to changes of the spin and path Bell states. The operation $U$ is also based on Hu’s work \cite{13} and composed of four BDs and ten LCs. The optical axes of LCs are all set at $45^\circ$. LC2, LC4, LC5, LC7, LC9 and LC10 are set to introduce 0 phase, while the others are set to introduce $\pi$ phase. Together with the first four LCs, Alice can encode $\left| {{\Phi _{58}}} \right\rangle $ for SDC. Any DP has no effect on spin while any LC has no effect on OAM. Here YES (NO) means the linear optical devices are (are not) in the optical path. }
\end{figure}

\begin{figure*}[htb]
\includegraphics[width=0.9\textwidth]{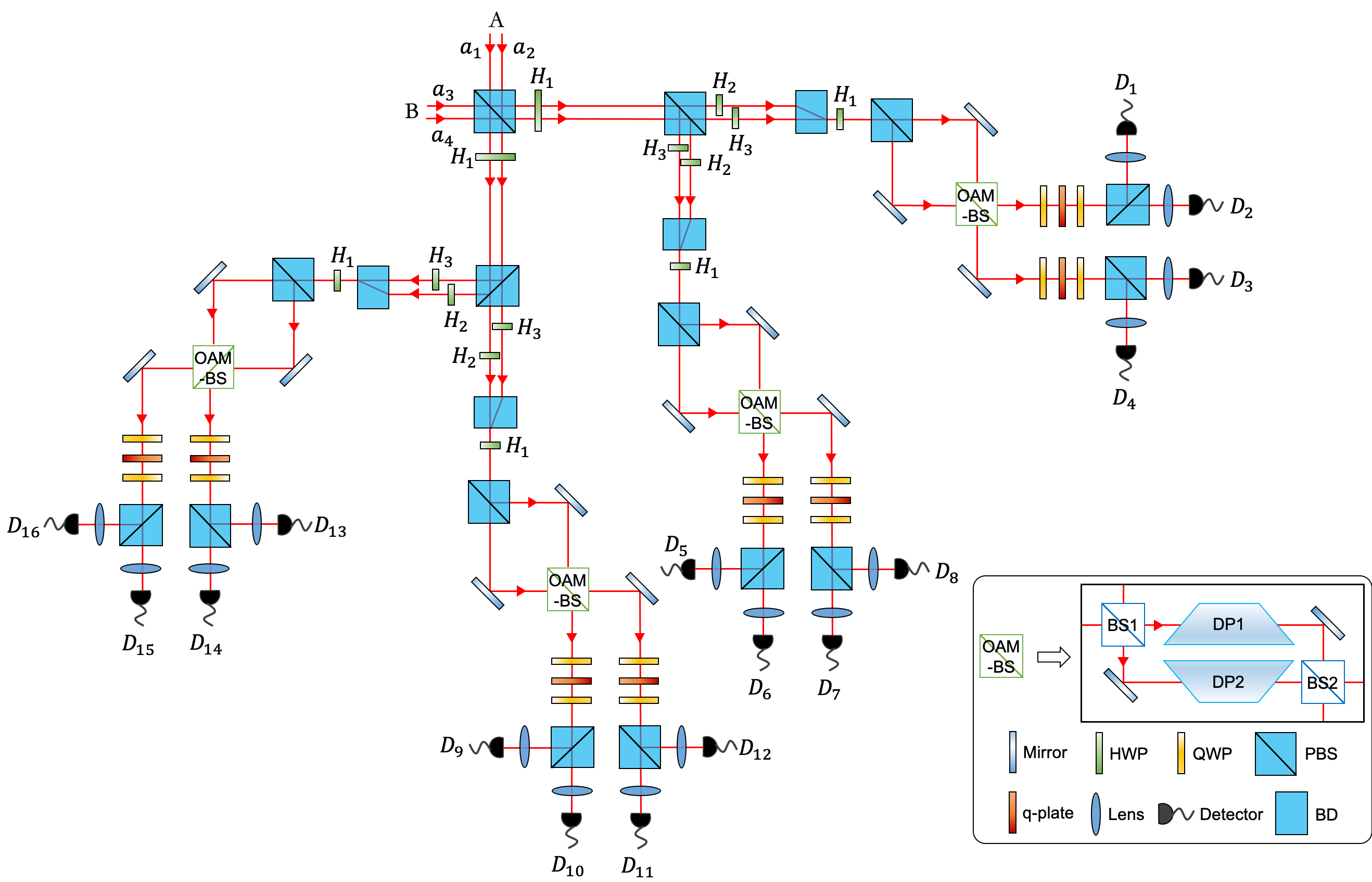}
\caption{\label{fig3} \textbf{Measurement of two-photon eight-dimensional Bell states.} This experimental setup combines Hu’s \cite{13} and Kong’s \cite{22} work and can be divided into two parts by four OAM-BS \cite{14,22}. OAM-BS is like a modified MZ interferometer composed of two BSs and two DPs. The relative orientation angle between DP1 and DP2 is set to be $45^\circ$. Any OAM-BS has no influence on spin of photon. The part before OAM-BSs is used to decode spin-path Bell states. HWPs are set at special degrees to rotate the polarization of photons or compensate the optical path difference. Here, H1 is set at $22.5^\circ$, H2 is set at $0^\circ$ and H3 is set at $45^\circ$. The part after OAM-BS is used to decode OAM Bell states. A q-plate sandwiched by two QWPs and a PBS are used together to project the state into the fundamental Gauss mode, which is collected by detectors.}
\end{figure*}

\section{\label{sec:level3}EXPERIMENT}
In the theory section, we have analyzed the optimal BSM schemes in different quantum systems. For a two-particle eight-dimensional hyperentangled system, $N_1=15$. Kong \emph{et~al}. experimentally achieved the complete distinction of eight Bell-like states using linear optical elements \cite{22}, which did not achieve the distinguishing limit of this system.
Thus, in this section, we would present an eight-dimensional BSM scheme theoretically based on two-photon OAM-spin-path hyperentanglement \cite{31} composed of two-dimensional OAM ($ \pm 1$ order), spin and path, which distinguishes 15 Bell-like states completely.

\begin{figure*}[htb]
\includegraphics[width=0.9\textwidth]{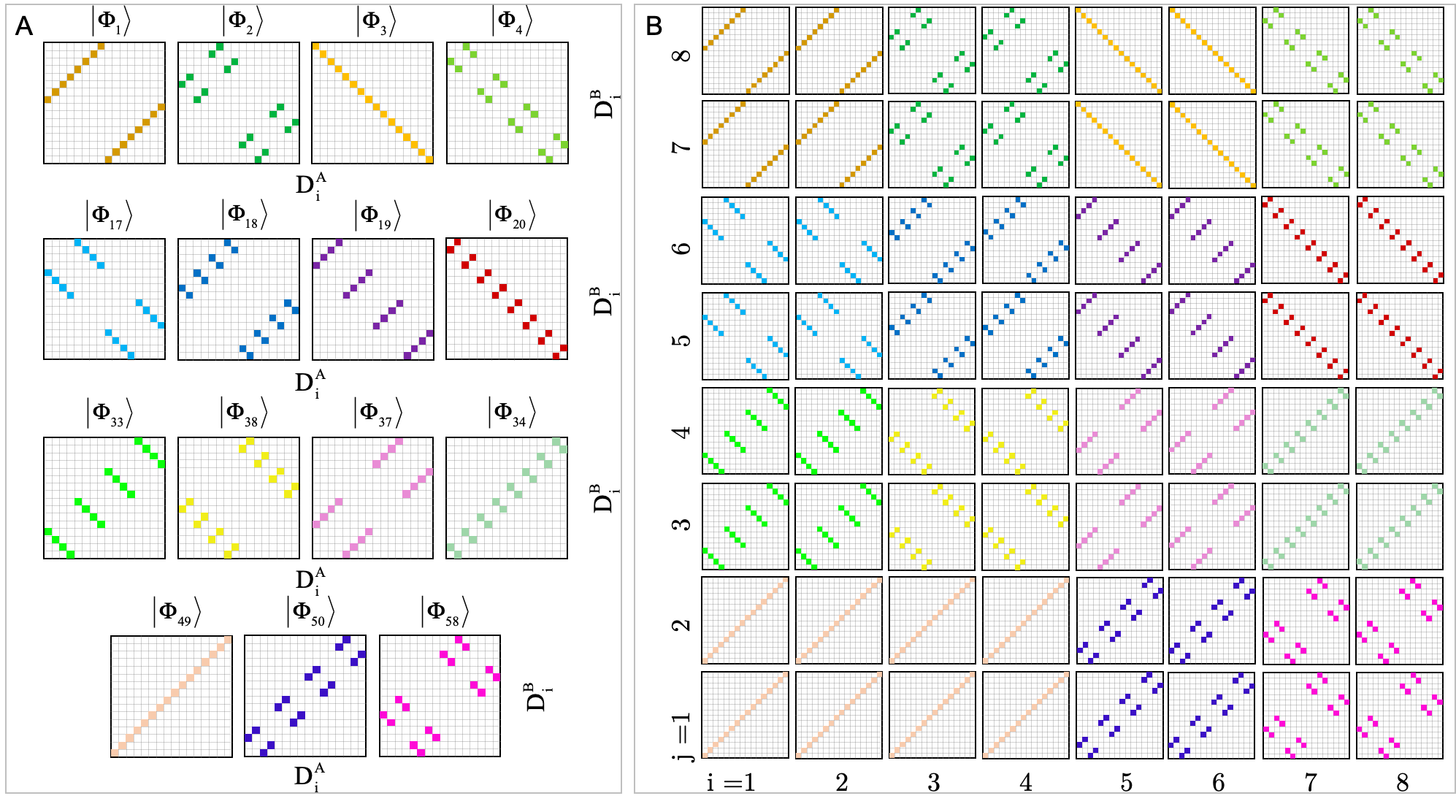}
\caption{\label{fig4} \textbf{Theoretical results of coincidence measurement for 15 and all 64 Bell-like states. } (A) Simulated coincidence measurement results of 15 Bell-like states with possible detector response from $D_{\rm{i}}^{\rm{A}}$ and $D_{\rm{i}}^{\rm{B}}$ ($i=1,2,...,16$). e.\ g.\ the corresponding coincidence between detectors for $\left| {{\Phi _1}} \right\rangle$ is ${D_{1,9}}+{D_{2,10}}+{D_{3,11}}+{D_{4,12}}+{D_{5,13}}+{D_{6,14}}+{D_{7,15}}+{D_{8,16}}$. The colored small squares (empties) mean there are (are not) coincidence counts. (B) For all the 15 classes of Bell-like states, there are 14 classes composed of 4 Bell-like states while only one class composed of 8 Bell-like states.}
\end{figure*}

First, 15 two-photon Bell-like states we are to focus on can be formulated as
\begin{eqnarray}\label{eq8}
    \left| {{\Phi _{1,2}}} \right\rangle  = {{\left( {\left| {\Phi _{{\rm{spin}}}^ + } \right\rangle  \otimes \left| {\Theta _{{\rm{path}}}^ \pm } \right\rangle  \otimes \left| {\Psi _{{\rm{OAM}}}^ + } \right\rangle } \right)} \mathord{\left/
 {\vphantom {{\left( {\left| {\Phi _{{\rm{spin}}}^ + } \right\rangle  \otimes \left| {\Theta _{{\rm{path}}}^ \pm } \right\rangle  \otimes \left| {\Psi _{{\rm{OAM}}}^ + } \right\rangle } \right)} {\sqrt 2 }}} \right.
 \kern-\nulldelimiterspace} {\sqrt 2 }},\nonumber \\
 \left| {{\Phi _{3,4}}} \right\rangle  = {{\left( {\left| {\Phi _{{\rm{spin}}}^ - } \right\rangle  \otimes \left| {\Theta _{{\rm{path}}}^ \pm } \right\rangle  \otimes \left| {\Psi _{{\rm{OAM}}}^ + } \right\rangle } \right)} \mathord{\left/
 {\vphantom {{\left( {\left| {\Phi _{{\rm{spin}}}^ - } \right\rangle  \otimes \left| {\Theta _{{\rm{path}}}^ \pm } \right\rangle  \otimes \left| {\Psi _{{\rm{OAM}}}^ + } \right\rangle } \right)} {\sqrt 2 }}} \right.
 \kern-\nulldelimiterspace} {\sqrt 2 }},\nonumber \\
 \left| {{\Phi _{17,18}}} \right\rangle  = {{\left( {\left| {\Phi _{{\rm{spin}}}^ + } \right\rangle  \otimes \left| {\Theta _{{\rm{path}}}^ \pm } \right\rangle  \otimes \left| {\Phi _{{\rm{OAM}}}^ + } \right\rangle } \right)} \mathord{\left/
 {\vphantom {{\left( {\left| {\Phi _{{\rm{spin}}}^ + } \right\rangle  \otimes \left| {\Theta _{{\rm{path}}}^ \pm } \right\rangle  \otimes \left| {\Phi _{{\rm{OAM}}}^ + } \right\rangle } \right)} {\sqrt 2 }}} \right.
 \kern-\nulldelimiterspace} {\sqrt 2 }},\nonumber \\
 \left| {{\Phi _{19,20}}} \right\rangle  = {{\left( {\left| {\Phi _{{\rm{spin}}}^ - } \right\rangle  \otimes \left| {\Theta _{{\rm{path}}}^ \pm } \right\rangle  \otimes \left| {\Phi _{{\rm{OAM}}}^ + } \right\rangle } \right)} \mathord{\left/
 {\vphantom {{\left( {\left| {\Phi _{{\rm{spin}}}^ - } \right\rangle  \otimes \left| {\Theta _{{\rm{path}}}^ \pm } \right\rangle  \otimes \left| {\Phi _{{\rm{OAM}}}^ + } \right\rangle } \right)} {\sqrt 2 }}} \right.
 \kern-\nulldelimiterspace} {\sqrt 2 }}, \\
 \left| {{\Phi _{33,34}}} \right\rangle  = {{\left( {\left| {\Theta _{{\rm{spin}}}^ + } \right\rangle  \otimes \left| {\Theta _{{\rm{path}}}^ \pm } \right\rangle  \otimes \left| {\Psi _{{\rm{OAM}}}^ + } \right\rangle } \right)} \mathord{\left/
 {\vphantom {{\left( {\left| {\Theta _{{\rm{spin}}}^ + } \right\rangle  \otimes \left| {\Theta _{{\rm{path}}}^ \pm } \right\rangle  \otimes \left| {\Psi _{{\rm{OAM}}}^ + } \right\rangle } \right)} {\sqrt 2 }}} \right.
 \kern-\nulldelimiterspace} {\sqrt 2 }},\nonumber \\
 \left| {{\Phi _{37,38}}} \right\rangle  = {{\left( {\left| {\Theta _{{\rm{spin}}}^ + } \right\rangle  \otimes \left| {\Theta _{{\rm{path}}}^ \pm } \right\rangle  \otimes \left| {\Psi _{{\rm{OAM}}}^ - } \right\rangle } \right)} \mathord{\left/
 {\vphantom {{\left( {\left| {\Theta _{{\rm{spin}}}^ + } \right\rangle  \otimes \left| {\Theta _{{\rm{path}}}^ \pm } \right\rangle  \otimes \left| {\Psi _{{\rm{OAM}}}^ - } \right\rangle } \right)} {\sqrt 2 }}} \right.
 \kern-\nulldelimiterspace} {\sqrt 2 }},\nonumber \\
 \left| {{\Phi _{49,50}}} \right\rangle  = {{\left( {\left| {\Theta _{{\rm{spin}}}^ + } \right\rangle  \otimes \left| {\Theta _{{\rm{path}}}^ \pm } \right\rangle  \otimes \left| {\Phi _{{\rm{OAM}}}^ + } \right\rangle } \right)} \mathord{\left/
 {\vphantom {{\left( {\left| {\Theta _{{\rm{spin}}}^ + } \right\rangle  \otimes \left| {\Theta _{{\rm{path}}}^ \pm } \right\rangle  \otimes \left| {\Phi _{{\rm{OAM}}}^ + } \right\rangle } \right)} {\sqrt 2 }}} \right.
 \kern-\nulldelimiterspace} {\sqrt 2 }},\nonumber \\
 \left| {{\Phi _{58}}} \right\rangle  = {{\left( {\left| {\Theta _{{\rm{spin}}}^ + } \right\rangle  \otimes \left| {\Psi _{{\rm{path}}}^ - } \right\rangle  \otimes \left| {\Phi _{{\rm{OAM}}}^ + } \right\rangle } \right)} \mathord{\left/
 {\vphantom {{\left( {\left| {\Theta _{{\rm{spin}}}^ + } \right\rangle  \otimes \left| {\Psi _{{\rm{path}}}^ - } \right\rangle  \otimes \left| {\Phi _{{\rm{OAM}}}^ + } \right\rangle } \right)} {\sqrt 2 }}} \right.
 \kern-\nulldelimiterspace} {\sqrt 2 }}.\nonumber
\end{eqnarray}

Here spin, path and OAM Bell states are defined respectively as
\begin{eqnarray} \label{eq9}
    \left| {\Phi _{{\rm{spin}}}^ \pm } \right\rangle  &=& {{\left( {{{\left| H \right\rangle }_A}{{\left| H \right\rangle }_B} \pm {{\left| V \right\rangle }_A}{{\left| V \right\rangle }_B}} \right)} \mathord{\left/
 {\vphantom {{\left( {{{\left| H \right\rangle }_A}{{\left| H \right\rangle }_B} \pm {{\left| V \right\rangle }_A}{{\left| V \right\rangle }_B}} \right)} {\sqrt 2 }}} \right.
 \kern-\nulldelimiterspace} {\sqrt 2 }},\nonumber \\
 \left| {\Theta _{{\rm{spin}}}^ \pm } \right\rangle  &=& {{\left( {{{\left| H \right\rangle }_A}{{\left| V \right\rangle }_B} \pm {{\left| V \right\rangle }_A}{{\left| H \right\rangle }_B}} \right)} \mathord{\left/
 {\vphantom {{\left( {{{\left| H \right\rangle }_A}{{\left| V \right\rangle }_B} \pm {{\left| V \right\rangle }_A}{{\left| H \right\rangle }_B}} \right)} {\sqrt 2 }}} \right.
 \kern-\nulldelimiterspace} {\sqrt 2 }},\nonumber \\
 \left| {\Theta _{{\rm{path}}}^ \pm } \right\rangle  &=& {{\left( {{{\left| {{a_1}} \right\rangle }_A}{{\left| {{a_3}} \right\rangle }_B} \pm {{\left| {{a_2}} \right\rangle }_A}{{\left| {{a_4}} \right\rangle }_B}} \right)} \mathord{\left/
 {\vphantom {{\left( {{{\left| {{a_1}} \right\rangle }_A}{{\left| {{a_3}} \right\rangle }_B} \pm {{\left| {{a_2}} \right\rangle }_A}{{\left| {{a_4}} \right\rangle }_B}} \right)} {\sqrt 2 }}} \right.
 \kern-\nulldelimiterspace} {\sqrt 2 }},\\
 \left| {\Psi _{{\rm{path}}}^ \pm } \right\rangle  &=& {{\left( {{{\left| {{a_1}} \right\rangle }_A}{{\left| {{a_4}} \right\rangle }_B} \pm {{\left| {{a_2}} \right\rangle }_A}{{\left| {{a_3}} \right\rangle }_B}} \right)} \mathord{\left/
 {\vphantom {{\left( {{{\left| {{a_1}} \right\rangle }_A}{{\left| {{a_4}} \right\rangle }_B} \pm {{\left| {{a_2}} \right\rangle }_A}{{\left| {{a_3}} \right\rangle }_B}} \right)} {\sqrt 2 }}} \right.
 \kern-\nulldelimiterspace} {\sqrt 2 }},\nonumber \\
 \left| {\Psi _{{\rm{OAM}}}^ \pm } \right\rangle  &=& {{\left( {{{\left| { + 1} \right\rangle }_A}{{\left| { + 1} \right\rangle }_B} \pm {{\left| { - 1} \right\rangle }_A}{{\left| { - 1} \right\rangle }_B}} \right)} \mathord{\left/
 {\vphantom {{\left( {{{\left| { + 1} \right\rangle }_A}{{\left| { + 1} \right\rangle }_B} \pm {{\left| { - 1} \right\rangle }_A}{{\left| { - 1} \right\rangle }_B}} \right)} {\sqrt 2 }}} \right.
 \kern-\nulldelimiterspace} {\sqrt 2 }},\nonumber \\
 \left| {\Phi _{{\rm{OAM}}}^ \pm } \right\rangle  &=& {{\left( {{{\left| { + 1} \right\rangle }_A}{{\left| { - 1} \right\rangle }_B} \pm {{\left| { + 1} \right\rangle }_A}{{\left| { - 1} \right\rangle }_B}} \right)} \mathord{\left/
 {\vphantom {{\left( {{{\left| { + 1} \right\rangle }_A}{{\left| { - 1} \right\rangle }_B} \pm {{\left| { + 1} \right\rangle }_A}{{\left| { - 1} \right\rangle }_B}} \right)} {\sqrt 2 }}} \right.
 \kern-\nulldelimiterspace} {\sqrt 2 }}.\nonumber
\end{eqnarray}
Here $H$($V$) represents the horizontal (vertical) polarization, $a_1$ and $a_2$ ($a_3$ and $a_4$) represents two different paths for photon A (B), $\left| { + 1} \right\rangle$ ($\left| { - 1} \right\rangle $) denotes a state of photon with an OAM of $ + \hbar $ ($-\hbar$). All the 15 Bell-like states showed in Eq.\eqref{eq8} are in identical eight-dimensional Hilbert space.
We perform unitary operation ${U^O} \otimes {U^S} \otimes {U^P}$ on the state $\left| {{\Phi _1}} \right\rangle $ of photon A to prepare the 15 Bell-like states described in Eq. \eqref{eq8} (Fig.\ \ref{fig2})

In order to unambiguously distinguish 15 Bell-like states mentioned in Eq.\ \eqref{eq8}, we design a theoretical optimal BSM scheme based on Hu’s \cite{13} and Kong’s work \cite{22} (Fig.\ \ref{fig3}). Only linear optical elements such as PBS, HWP, QWP, q-plate, DP (Dove prism) BS (beam splitter) and BD (beam displacer) are used.

According to this experimental setup (Fig.\ \ref{fig3}), we can derive the coincidence between detectors for 15 Bell-like states in Eq.\ \eqref{eq8} (Fig.\ \ref{fig4}A) and classify all 64 Bell-like states into 15 different classes (Fig.\ \ref{fig4}B). The results show that any one of the 15 Bell-like states have a unique pattern of coincidence between detectors. That is to say, all these Bell-like states can be distinguished completely. This result corresponds to the theory in the previous section, since the independent constraint equation here is ${P_{12}}\left| {{D_1}} \right\rangle \left| {{D_8}} \right\rangle {\rm{ = }}0$. Moreover, it is worthy to note that $\left| {{\Phi _{49}}} \right\rangle$ has 16 two-fold coincidence between the same detector. Thus, one can distinguish 14 classes of Bell-like states without number-resolving detectors.

Here we want to further discuss the meaning of our experiment. Just as the experiment shows, if we want to realize an optimal BSM for two-photon systems experimentally, mixing particles in their channels is a must. For instance, in the decode (Bob) part in our experiment, the two laser beams go into the same PBS from different directions at first. This is the key to success. As a result, every single photon has chance to trigger every single detector. Thus, there are more possibilities for the coincidences between detectors. So as to the projective basis in the experiment. So, more classes of Bell states can be distinguished in the end. According to Pisenti’s results \cite{20}, when n particles are mixed, the upper limit is $nD$ when we ignore the constraint part. If not mixed, the same bound can reduce to only $D$. Take Kong’s \cite{22} and Hu’s \cite{13} work for examples. Hu’s work can realize an optimal BSM theoretically, and the setup has the feature on PBS corresponding to the discussion above. Compared with Hu’s setup, Kong’s setup split the pump into two beams and each of them passes through a Bell-like state analyzer separately. Therefore, it lacks the coincidences between the detectors from the same Bell-like state analyzer and only 8 Bell-like states can be distinguished from 64 ones.

\section{\label{sec:level4}SUMMARY}

In summary, we have obtained the distinguishing limit $N$ for any $nD$ entangled system. For those linear-optics circuits with $G_1$, we have proved that ${N_1}=nD-(n-1)$ , which can be applied to both bosons’ and fermions’ cases. It is worth mentioning that being different from multi-photon systems, ${N_1}=2D-1$ is the upper limit for two-photon systems. For those circuits with $G_2$, we infer that ${N_2} \ge {D^{n - 1}}$ due to the symmetry of Bell states. Thus, for those $n$-photon entangled systems with $n>2$, one can at least distinguish $D^{n-1}$ Bell states with optimal BSM schemes. We also verify the result for $N_1$ using LL criterion. Particularly, since for those circuits with $G_1$ the norm Eq.\ \eqref{eq13} is always positive, we have proved that Eq.\ \eqref{eq2} is a sufficient and necessary condition to distinguish two states here. This can help largely reduce and simplify the computation. Based on this result, we can choose a better entangled system to realize BSM effectively and check quickly if any given BSM scheme has reached the limit $N$. Moreover, we have demonstrated the BSM scheme theoretically for two-photon eight-dimensional hyperentanglement using spin, path and the first order of OAM, based on Kong’s \cite{22} and Hu’s \cite{13} work. The highlight is this scheme helps to achieve the CC’s upper bound ${\log _2}15$  by dividing 15 different classes out of 64 Bell states.

\section{ACKNOWLEDGEMENT}

Supported by the National Key R\&D Program of China under Grant Nos 2017YFA0303800 and 2017YFA0303700, the National Natural  Science Foundation of China under Grant Nos 11534006, 11674184 and 11774183 the Natural Science Foundation of Tianjin under Grant No 16JCZDJC31300, and the Collaborative Innovation Center of Extreme Optics.

\nocite{*}

\bibliography{main}

\end{document}